\documentclass[journal]{IEEEtran}
\usepackage{multirow}
\usepackage{graphicx}
\graphicspath{{images/}}
\usepackage{xcolor}
\usepackage{ulem}
\usepackage{threeparttable}
\usepackage{tipa}

\usepackage{CJK}
\usepackage{float} 
\usepackage{cite}
\usepackage[numbers,sort&compress]{natbib}
\usepackage{amsmath}
\usepackage{amssymb}
\usepackage{algorithm}
\usepackage{algorithmic}
\usepackage{diagbox}
\usepackage{color}
\usepackage{array}
\usepackage{booktabs}

\usepackage{colortbl}

\ifCLASSINFOpdf
\else
\fi
\begin{document}
%
\title{Enhancing Segment-Based Speech Emotion Recognition by Deep Self-Learning}
%
%
%

\author{Shuiyang Mao, P. C. Ching, and Tan Lee
\thanks{S. Mao, P. C. Ching, and T. Lee are with the Department of Electronic Engineering, The Chinese University of Hong Kong, Hong Kong (e-mail: maoshuiyang@link.cuhk.edu.hk; pcching@ee.cuhk.edu.hk; tanlee@ee.cuhk.edu.hk).}
}

%
%

\markboth{Journal of \LaTeX\ Class Files,~Vol.~14, No.~8, March~2021}%
{Shell \MakeLowercase{\textit{et al.}}: Bare Demo of IEEEtran.cls for IEEE Journals}
%



\maketitle

\begin{abstract}
  Despite the widespread utilization of deep neural networks (DNNs) for speech emotion recognition (SER), they are severely restricted due to the paucity of labeled data for training. Recently, segment-based approaches for SER have been evolving, which train backbone networks on shorter segments instead of whole utterances, and thus naturally augments training examples without additional resources. However, one core challenge remains for segment-based approaches: most emotional corpora do not provide ground-truth labels at the segment level. To supervisely train a segment-based emotion model on such datasets, the most common way assigns each segment the corresponding utterance's emotion label. However, this practice typically introduces noisy (incorrect) labels as emotional information is not uniformly distributed across the whole utterance. On the other hand, DNNs have been shown to easily over-fit a dataset when being trained with noisy labels. To this end, this work proposes a simple and effective deep self-learning (DSL) framework, which comprises a procedure to progressively correct segment-level labels in an iterative learning manner. The DSL method produces dynamically-generated and soft emotion labels, leading to significant performance improvements. Experiments on three well-known emotional corpora demonstrate noticeable gains using the proposed method.
\end{abstract}

\begin{IEEEkeywords}
  Segment-based speech emotion recognition, learning with noisy labels, deep self-learning, soft labeling.
\end{IEEEkeywords}

%
\IEEEpeerreviewmaketitle

\section{Introduction}
%
%
%
%
\IEEEPARstart{S}{peech} emotion recognition (SER) aims at decoding emotional content from speech signals. It has constituted an active topic in the research area of human-machine interaction (HCI). In particular, monitoring call center services, detection of lies, and medical diagnoses are often considered promising application scenarios of SER.

Prior research in SER has primarily focused on the \textit{utterance}-based approach, where the backbone model is performed on the whole utterance. One major problem in the state-of-the-art utterance-based paradigm is the scarcity of the training data compared to model complexity. Popular emotional speech databases, such as the Emo-DB corpus \cite{burkhardt2005database} and the SAVEE database \cite{jackson2014surrey}, contain recordings of approximately one-hour speech only; the number of utterance-level feature vectors may not be sufficient for reliably estimating model parameters of a complex classifier such as deep neural networks (DNNs). An alternative is the \textit{segment}-based approach~\cite{shami2005segment, schuller2006timing, mower2009interpreting, mower2010framework, jeon2011sentence, han2014speech, satt2017efficient, mao2019deep, maoemotion}, in which the backbone model can be trained more reliably on a large number of individual segments, which naturally augments training examples without extra resources.

One major shortcoming remains for the segment-based approach: Most emotional speech databases do not provide labels at the segment level. To supervisely train an emotion segment model, the most common way is to assign the utterance-level label to each segment \cite{schuller2006timing, han2014speech, satt2017efficient, mao2019deep}. Such labeling practice, however, may introduce \textit{noisy} (incorrect) labels, since emotional information are not uniformly distributed over all positions of an utterance \cite{schuller2006timing, jeon2011sentence, rao2013emotion}. 

Another source of \textit{label noise} in the segment-based approach comes from the dataset itself. For instance, most \textit{acted} emotional corpora do not well enough simulate emotions naturally and clearly \cite{el2011survey}. In other words, the intended emotions are often not properly expressed. This is evidenced by the relatively poor recognition performance of human subjects; in \cite{nwe2003speech}, the reported human recognition rate is as low as about 65\%. Also, the problem of \textit{elicited} and \textit{natural} emotional corpora lies in the label annotation; their emotion labels are often annotated based on human annotators' perception of emotions. However, in many cases, one may not clearly distinguish one emotion from another, which may introduce errors in the annotation process. Based on the above discussion, we can see that the labels provided by the dataset itself inevitably contain label noise due to subjective expression (for acted emotional corpora) or subjective perception (for elicited or natural emotional corpora) of emotions. Consequently, the segments that merely inherit their labels from corresponding utterances may also inherit the same inferiority of subjective labeling of emotional speech. 

On the other hand, it has been widely reported that label noise significantly degrades the generalization performance of complex models such as DNNs since they easily over-fit the label noise \cite{nettleton2010study, xiao2015learning, han2019deep}. Therefore, limiting the negative influence of label noise is of great practical importance. In computer vision, many efforts have been made to improve the robustness of DNNs trained on noisy labels \cite{frenay2013classification, sukhbaatar2014training, patrini2017making, hendrycks2018using, zhang2018generalized, veit2017learning}. However, similar efforts have not been attempted in the segment-based SER systems to our best knowledge. Herein, it is our goal to train a robust emotion segment model on the noisy labeled segments. Specifically, we introduce deep self-learning (DSL), a solution that uses a DNN model and the data to correct segment-level labels during iterative training. This will be detailed in Section~\ref{sec:dsl}.

Another critical issue in the SER task is that human emotions are inherently impure. When designing systems to recognize human emotions, the emotional impurity must be considered. However, most of the current methods rest on the consensus, e.\,g., one single hard label for an utterance (\textit{hard labeling}). This labeling principle imposes specific challenges on SER related tasks, e.\,g., incomplete labeling. For instance, frustration can overlap with other emotion categories ranging from anger to neutral and sadness \cite{mower2009interpreting, busso2008iemocap}. SER systems designed to output one hard label for each speech utterance (or segment) may perform poorly if the target expression cannot be well captured by a single emotion label \cite{mower2010framework}. 

A natural solution to the above problem is to perform \textit{soft labeling}, which characterizes expressions as complex mixtures of possible emotions, rather than one-hot hard assignments. This work explores the soft labeling approach in the DSL framework. Specifically, three soft labeling-based strategies are proposed to construct training targets during iterative learning, i.\,e., basic dynamic soft labeling (BDSL), weighted dynamic soft labeling (WDSL), and global soft labeling (GSL). For comparison, we also investigate a hard labeling-based method called dynamic hard labeling (DHL). In our experiments, the soft labeling-based methods outperform the hard labeling-based one overall. In particular, the WDSL method achieves the best performance.

The contributions of this work are summarized as follows: (1) We propose an iterative learning framework DSL to improve the performance of segment-based SER task. The segments that inherit their labels from the corresponding utterance are treated as the ``\textit{noisy dataset}'', and the training of segment-based emotion model on it is framed as the ``\textit{learning with noisy labels}'' problem. (2) We empirically explore and compare various deep convolutional neural network (DCNN) architectures as the backbone for the SER task. (3) We demonstrate the capability of a network to improve accuracy by training against labels generated by another network of the same architecture. (4) Extensive experiments on three popular emotional corpora are conducted, and the experimental results consistently validate the effectiveness of our methods. In particular, when EfficientNet-B0 \cite{tan2019efficientnet} is used as the backbone model, our method improves recognition rate on the CASIA corpus \cite{tao2008design} from \(95.17\%\) to \(96.82\%\) (WA and UA), the Emo-DB database \cite{burkhardt2005database} from \(83.36\%\) to \(92.90\%\) (WA) and \(82.54\%\) to \(93.02\%\) (UA), and the SAVEE database \cite{jackson2014surrey} from \(73.75\%\) to \(86.46\%\) (WA) and \(72.26\%\) to \(85.71\%\) (UA), achieving new state-of-the-art performance on all three databases. 



\begin{figure}[t]
  \centering
  \includegraphics[width=\linewidth]{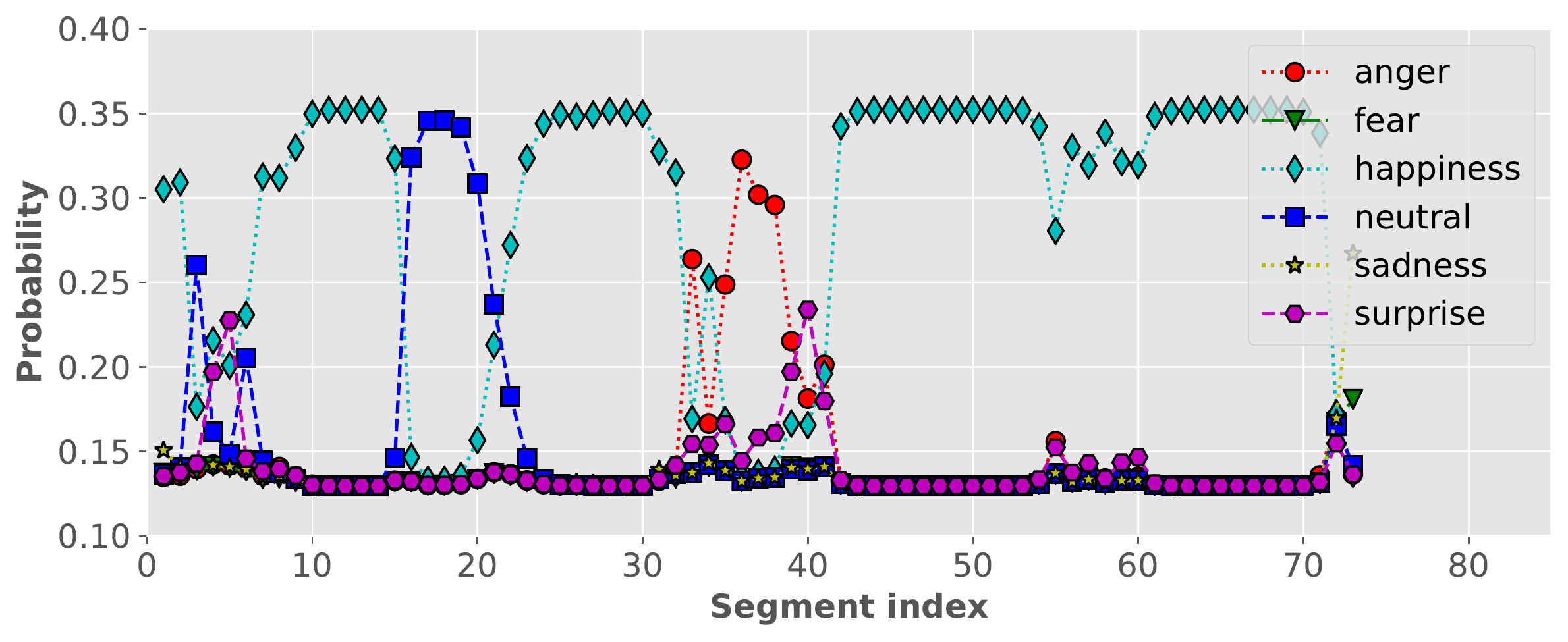}
  \caption{An example of segment-level label predictions across the audio file ``Happy\_liuchanhg\_382.wav'' in the CASIA corpus. A VGG network was used as the backbone model.}
  \label{fig:prop}
\end{figure}

\begin{figure*}[!htbp]
  \centering
  \includegraphics[width=0.8\linewidth]{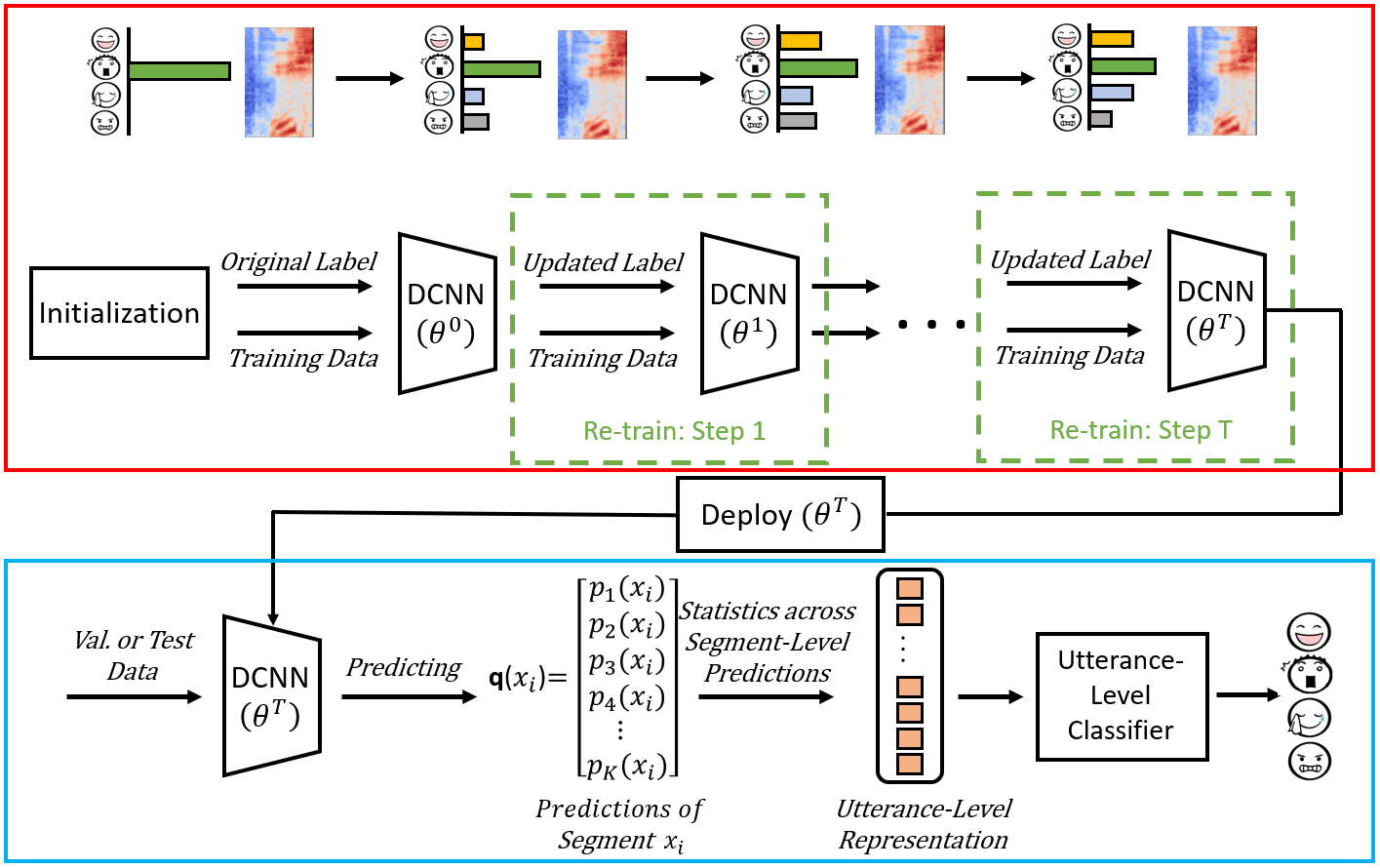}
  \caption{Illustration of our system. The red block shows the proposed DSL framework, in which we train the backbone DCNN model and correct labels for each segment alternatively. The blue block shows that we use the segment-level predictions generated by the latest backbone DCNN to construct utterance-level representations for utterance-level classification.}
  \label{fig:framework}
\end{figure*}

\section{Related Work}
\label{sec:related-work}
\textbf{Data Augmentation in SER:} The scarcity of training data for emotional speech seriously limits the generalization performance of current SER systems. Data augmentation appears to be the most common way to address this problem. For instance, Etienne et al. augmented emotional speech data through a combination of oversampling and vocal tract length perturbation (VTLP) \cite{etienne2018cnn+}. \cite{zhu2017data, yi2019adversarial, chatziagapi2019data} employed generative adversarial networks (GANs) to generate training samples. Motivated by related works in computer vision, transfer learning has also been applied to SER tasks \cite{deng2013sparse, gideon2017progressive}. This work adopts a segment-based framework in which the backbone model is trained reliably on a large number of individual segments. This naturally augments the training vectors without extra resources or an accessorial network.

\textbf{Learning with Noisy Labels:} When training data are corrupted by label noise, an obvious solution is to filter out the erroneous labels themselves. Indeed, many methods have been proposed to find and remove mislabeled instances, with different degrees of success \cite{frenay2013classification}. However, a robust mechanism to identify and remove erroneous labels from the label set remains lacking. The \textit{noise transition matrix} has also been widely utilized to characterize the transition probability between the observed (noisy) label and the latent true label \cite{sukhbaatar2014training, patrini2017making, hendrycks2018using}. However, the above methods follow a simple assumption that the transition probability is independent of individual instances, which often does not hold in real-world noisy datasets and thus limits their performance. Furthermore, various noise-robust loss functions \cite{zhang2018generalized, veit2017learning} have been explored to combat label noise. However, most of these functions either have an assumption similar to the above transition matrix-based approaches or require extra clean samples, limiting their applications in practice. In contrast, the DSL framework adopted in this work does not rely on any assumption about the distribution of label noise, making it feasible for large-scale real-world scenarios. It also does not require additional clean samples or extra supervision, providing a simple and effective ``\textit{learning with noisy labels}'' scheme.   

\textbf{Soft Labeling in SER:} Soft labeling approaches have recently received increased focus for modeling emotional ambiguity and impurity. For example, Lotfian et al. proposed a novel probabilistic approach for soft labeling of emotional speech \cite{lotfian2017formulating}. Ando et al. developed a DNN-based model trained with soft emotion targets as ground truth to better capture emotional ambiguity \cite{ando2018soft}. Kim et al. utilized soft labeling and cross-entropy to compare emotion label distributions made by humans and machines \cite{kim2018human}. All of the above works performed soft labeling on the whole utterance. In contrast, this work aims at the individual segments, of which the emotional expressions might be more ambiguous than that of the longer utterances.

\section{Segment-based SER}
\label{sec:method}
The segment-based approach for SER has existed in the research community for some years \cite{schuller2006timing, mower2009interpreting, mower2010framework, jeon2011sentence, han2014speech, satt2017efficient, mao2019deep}. Empirical comparisons between segment-based processing and traditional utterance-based approach in prior studies have demonstrated the superiority of the former \cite{schuller2006timing, jeon2011sentence, han2014speech, satt2017efficient, mao2019deep}. The key idea of the segment-based approach for SER is first to obtain classification results of the emotional state at segment level. Then, these segment-level decisions are aggregated to form utterance-level representations for utterance-level classification.

However, since detailed annotation of the speech utterance frequently constitutes an ambiguous and expensive task, most emotional speech databases do not provide segment-level labels; instead, we only have labels for the whole utterance. Consequently, the segment-based method must address the problem of how to infer an emotion segment model without access to a training set of labeled segments. To circumvent this problem, a common practice is to follow the most straightforward approach, i.\,e., each segment inherits the label of corresponding utterance \cite{han2014speech, satt2017efficient, mao2019deep}. A segment-based emotion model is then trained on the resulting dataset. The trained model aims to output a probability prediction \(\textbf{q}\) over all emotional states for each segment:
\begin{align}
  \textbf{q}(\mathbf{x}_i) = [p_1(\mathbf{x}_i),\ p_2(\mathbf{x}_i),\ \cdots,\ p_K(\mathbf{x}_i)]^T \in \mathbb{R}^{K\times 1}
  \label{eq1}
\end{align}
where \(\mathbf{x}_i\) represents a specific segment; and \(K\) denotes the number of emotion categories. 

Figure~\ref{fig:prop} presents an example of segment-level emotion predictions across the audio file ``Happy\_liuchanhg\_382.wav'' in the CASIA corpus, where a VGG network \cite{simonyan2014very} was used as the backbone model. It can be observed that: (1) The probability distribution of each segment changes over the whole utterance. (2) Although most segments convey information that conforms to the corresponding utterance, some segments are more related to other classes, introducing label noise and impeding model generalization performance. If we can correct those noisy (incorrect) labels -- a topic that has not yet been touched in the segment-based SER to our best knowledge, system performance might be improved.

Inspired by recent successes of self-learning in dealing with noisy labeled images in computer vision \cite{bagherinezhad2018label, tanaka2018joint, han2019deep}, here, we propose an iterative learning scheme with DCNN as the backbone model, which we call \textit{deep self-learning} (DSL) throughout this paper. We treat those individual segments that inherit their labels from corresponding utterances as the \textit{noisy data}, and their labels as \textit{noisy labels}. We then formulate the training of the emotion segment model as a ``\textit{learning with noisy labels}'' problem, in which the model parameters and labels are jointly optimized. The following section specifies our approach.

\section{Our Approach}
\label{sec:dsl}

\subsection{Problem Formulation}
In a supervised \(K\)-class classification problem setting, we have a dataset of \(N\) samples: \(\mathcal{D}= \{\mathbf{X}, \mathbf{Y}\} = \{(\mathbf{x}_1, \mathbf{y}_1), \dots, (\mathbf{x}_N, \mathbf{y}_N)\}\), and \(\mathbf{y}_i \in \{0, 1\}^K\) is the one-hot vector representation of the class label corresponding to the input feature \(\mathbf{x}_i\). Generally, the optimization problem is formulated as follows:
\begin{align}
  \mathop{\textnormal{min}} \limits_{\theta}\mathcal{L}(\theta | \mathbf{X}, \mathbf{Y})
  \label{clean_loss}
\end{align}
where \(\theta\) denotes the model parameters to be learned; and \(\mathcal{L}\) represents a loss objective function, such as the cross entropy. Eq.~(\ref{clean_loss}) works well on clean labels. However, when \(\mathbf{Y}\) contains noise (erroneous labels), the solution of Eq.~(\ref{clean_loss}) would be sub-optimal, limiting the generalization performance of the derived models.

In this work, we first assign each segment the emotion label of the corresponding utterance. This labeling procedure results in a dataset with noisy labels. To attain the latent true label \(\hat{\mathbf{Y}}\) from the initial set of segments with noisy labels and thus train a robust emotion segment model, we propose \textit{deep self-learning} (DSL), an effective framework to jointly optimize the model parameters and labels in an iterative learning manner. Our optimization problem can be formulated as follows:
\begin{align}
  \mathop{\textnormal{min}} \limits_{\theta, \hat{\mathbf{Y}}}\mathcal{L}(\theta, \hat{\mathbf{Y}} | \mathbf{X}, \mathbf{Y})
  \label{eq2}
\end{align}
where the network parameters and the class labels are updated alternatively. The details are described in the following subsections.

\subsection{DSL Framework}

\subsubsection{Pipeline}
Figure~\ref{fig:framework} illustrates a schematic diagram of the proposed method. We use deep convolutional neural networks (DCNN) as the backbone model trained on log Mel filterbanks of individual segments. The network training and label update proceed alternatively until the training converges. The segment-level predictions generated by the latest backbone DCNN are used to construct utterance-level feature vectors for utterance-level classification.

\subsubsection{Iterative Self-Learning}
The first backbone DCNN \(C_{\theta_{0}}\) is trained against the original noisy label. It takes individual segments, e.\,g., \(\mathbf{x}_i\) as input and produces corresponding label predictions \(\textbf{q}(\theta_{0}, \mathbf{x}_i)\). The second backbone DCNN \(C_{\theta_{1}}\) is trained over the same input, e.\,g., \(\mathbf{x}_i\), but uses updated labels \(\textnormal{g}(\textbf{q}(\theta_{0}, \mathbf{x}_i))\), where \(\textnormal{g}(.)\) represents the update rule of labels to construct training targets for training the next backbone DCNN. Details of the update rule of labels will be described in the next subsection. Once \(C_{\theta_{1}}\) is trained, we can similarly use the updated labels \(\textnormal{g}(\textbf{q}(\theta_{1}, \mathbf{x}_i))\) to train a subsequent network \(C_{\theta_{2}}\), etc. Intuitively, as the backbone networks improve over iterations of re-training, the labels are progressively corrected. 

We train the first network \(C_{\theta_{0}}\) using conventional cross-entropy loss. We train each of the subsequent network \(C_{\theta_{t}}\) for \(t \geq 1\) by minimizing the KL-divergence between its output \(\textbf{q}(\theta_{t}, \mathbf{x}_i)\) and the updated labels \(\textnormal{g}(\textbf{q}(\theta_{t-1}, \mathbf{x}_i))\) from the backbone DCNN \(C_{\theta_{t-1}}\) for \(t \geq 1\) in the previous iteration. Our loss function is formulated by: 
\begin{equation} \label{eq:kl_loss}
\begin{aligned}
  \mathcal{L}_t = &-\sum_{i=1}^{n}\textnormal{g}(\textbf{q}(\theta_{t-1}, \mathbf{x}_i))\;\textnormal{log}(\frac{\textbf{q}(\theta_{t}, \mathbf{x}_i)}{\textnormal{g}(\textbf{q}(\theta_{t-1}, \mathbf{x}_i))})\\
                = &-\sum_{i=1}^{n}\textnormal{g}(\textbf{q}(\theta_{t-1}, \mathbf{x}_i))\;\textnormal{log}(\textbf{q}(\theta_{t}, \mathbf{x}_i))\\
                  &+ \sum_{i=1}^{n}\textnormal{g}(\textbf{q}(\theta_{t-1}, \mathbf{x}_i))\;\textnormal{log}(\textnormal{g}(\textbf{q}(\theta_{t-1}, \mathbf{x}_i)))
\end{aligned}
\end{equation}
where \(n\) is the mini-batch size. The second term in Eq. (\ref{eq:kl_loss}) is constant with respect to \(C_{\theta_{t}}\). We can thus remove it and minimize the cross-entropy loss instead:
\begin{equation} \label{eq:entropy}
\begin{aligned}
  \mathcal{L}_t = -\sum_{i=1}^{n}\textnormal{g}(\textbf{q}(\theta_{t-1}, \mathbf{x}_i))\;\textnormal{log}(\textbf{q}(\theta_{t}, \mathbf{x}_i))
\end{aligned}
\end{equation}

\subsubsection{Updating Labels}
In this work, four update rules of labels are investigated, of which three are soft labeling-based, namely, basic dynamic soft labeling (BDSL), weighted dynamic soft labeling (WDSL), and global soft labeling (GSL). The remaining one is hard labeling-based, i.\,e., dynamic hard labeling (DHL). The details are outlined as follows.

\begin{itemize}
  \item \textbf{Basic Dynamic Soft Labeling (BDSL).} For the BDSL method, we directly use the model predictions, e.\,g., \(\textbf{q}(\theta_{t-1}, \mathbf{x}_i)\), to construct the supervision signal for training the next model \(C_{\theta_{t}}\) as follows:
  \begin{equation} \label{eq:bdsl}
    \begin{aligned}
      \textnormal{g}^{BDSL}(\textbf{q}(\theta_{t-1}, \mathbf{x}_i)) = \textbf{q}(\theta_{t-1}, \mathbf{x}_i)
    \end{aligned}
  \end{equation}
  
  The resulting loss function is:
  \begin{equation} \label{eq:entropy-bdsl}
    \begin{aligned}
      \mathcal{L}^{BDSL}_t = -\sum_{i=1}^{n}\textbf{q}(\theta_{t-1}, \mathbf{x}_i)\;\textnormal{log}(\textbf{q}(\theta_{t}, \mathbf{x}_i))
    \end{aligned}
  \end{equation} 

  \item \textbf{Weighted Dynamic Soft Labeling (WDSL).} The WDSL method uses a convex combination of original noisy labels, e.\,g., \(\mathbf{y}_i\), and the model predictions to update training labels as follows:
  \begin{equation} \label{eq:wdsl}
    \begin{aligned}
      \textnormal{g}^{WDSL}(\textbf{q}(\theta_{t-1}, \mathbf{x}_i)) = (1 - \alpha)\,\textbf{q}(\theta_{t-1}, \mathbf{x}_i) + \alpha\,\mathbf{y}_i
    \end{aligned}
  \end{equation}
  where \(\alpha \in [0, 1]\) is a coefficient balancing the two terms. 

  The corresponding loss function for the WDSL method is:
  \begin{equation} \label{eq:entropy-wdsl}
    \begin{aligned}
      \mathcal{L}^{WDSL}_t = &- (1 - \alpha) \sum_{i=1}^{n}\textbf{q}(\theta_{t-1}, \mathbf{x}_i)\;\textnormal{log}(\textbf{q}(\theta_{t}, \mathbf{x}_i))\\
                             &- \alpha \sum_{i=1}^{n}\mathbf{y}_i\;\textnormal{log}(\textbf{q}(\theta_{t}, \mathbf{x}_i))         
    \end{aligned}
  \end{equation} 

  \item \textbf{Global Soft Labeling (GSL).} To generate a global soft label for a certain speech segment \(\mathbf{x}_i\), we first pass all segments within the corresponding utterance \(\mathcal{U}\) to the current trained backbone network, and the global soft label is then computed by averaging the network outputs across the whole utterance as follows:
  \begin{equation} \label{eq:gsl}
    \begin{aligned}
      \textnormal{g}^{GSL}_k(\textbf{q}(\theta_{t-1}, \mathbf{x}_i)) = \frac{1}{|\mathcal{U}|}\sum_{\mathbf{x}_i \in \mathcal{U}}p_{k}(\theta_{t-1}, \mathbf{x}_i)
    \end{aligned}
  \end{equation}
  where \(k\) denotes the \(k^{th}\) emotion category.

  The corresponding loss function is:
  \begin{equation} \label{eq:entropy-gsl}
    \begin{aligned}
      \mathcal{L}^{GSL}_t = -\sum_{i=1}^{n} \sum_{k=1}^{K} \textnormal{g}^{GSL}_k(\textbf{q}(\theta_{t-1}, \mathbf{x}_i))\;\textnormal{log}(\textbf{q}(\theta_{t}, \mathbf{x}_i))
    \end{aligned}
  \end{equation} 

  \item \textbf{Dynamic Hard Labeling (DHL).} For the DHL method, the one-hot hard label is assigned to each speech segment by choosing the most-likely category from the corresponding network output as follows:
  \begin{equation} \label{eq:dhl}
    \textnormal{g}^{HDL}_k(\textbf{q}(\theta_{t-1}, \mathbf{x}_i)) =
    \left\{
      \begin{array}{lr}
        1, \ \textnormal{if}\ k = \textnormal{argmax}_{k^{\prime}}p_{k^{\prime}}(\theta_{t-1}, \mathbf{x}_i)\\
        0, \ \textnormal{otherwise}
        \end{array}
    \right.
  \end{equation}
  where \(k^{(\prime)}\) denotes the \(k^{(\prime)th}\) emotion category.

  The corresponding loss function can be expressed as:
  \begin{equation} \label{eq:entropy-dhl}
    \begin{aligned}
      \mathcal{L}^{HDL}_t = -\sum_{i=1}^{n} \sum_{k=1}^{K} \textnormal{g}^{HDL}_k(\textbf{q}(\theta_{t-1}, \mathbf{x}_i))\;\textnormal{log}(\textbf{q}(\theta_{t}, \mathbf{x}_i))
    \end{aligned}
  \end{equation} 

\end{itemize}

\subsection{Utterance-level Emotion Classification}
Algorithm~\ref{algo: sl} illustrates the overall process of training a robust backbone DCNN on noisy labeled speech segments using the proposed DSL framework. Segment-level predictions produced by the latest backbone network are used for constructing feature vectors for utterance-level classification. Specifically, the utterance-level representations are computed from the statistics of the segment-level predictions:
\begin{align}
  &h_1^k = \frac{1}{|\mathcal{U}|}\sum\limits_{\mathbf{x}_i \in \mathcal{U}} p_k(\mathbf{x}_i)\\
  &h_2^k = \mathop\text{1 \%-tile}_{\mathbf{x}_i \in \mathcal{U}}\{p_k(\mathbf{x}_i)\}\\
  &h_3^k = \mathop\text{99 \%-tile}_{\mathbf{x}_i \in \mathcal{U}}\{p_k(\mathbf{x}_i)\}\\
  &h_{4-6}^k = \mathop\text{Quartiles 1-3}_{\mathbf{x}_i \in \mathcal{U}}\{p_k(\mathbf{x}_i)\}\\  
  &h_{7-8}^k = \frac{|p_k(\mathbf{x}_i)>\beta|}{|\mathcal{U}|}
  \label{eq2}
\end{align}
where \(p_k(\mathbf{x}_i)\) is the probability of the \(k^{th}\) emotion for a specific segment \(\mathbf{x}_i\); and \(\mathcal{U}\) denotes the set of all segments from a certain utterance. \(h^k_1\), \(h^k_2\), \(h^k_3\), and \(f^k_{4-6}\) are the arithmetic mean, percentile \(1\), percentile \(99\), quartiles \(1\)-\(3\) of segment-level probabilities of the \(k^{th}\) emotion across an utterance, respectively. In particular, percentile \(1\) and percentile \(99\) serve as a robust substitute for the minimum value and maximum value, respectively. The remaining two features, \(h^k_{7-8}\), correspond to the percentage of segments which have higher probabilities than a given threshold for the \(k^{th}\) emotion. \(h^k_{7-8}\) are not sensitive to the threshold \(\beta\), and we herein heuristically set \(\beta\) equal to \(0.2\) for \(h^k_{7}\) and \(0.3\) for \(h^k_{8}\). This step results in a feature vector with a dimension of \(8 \times K\) for each utterance. With this collection of utterance-level feature vectors, we can train a second relatively simple classifier, i.\,e., random forest (RF) \cite{liaw2002classification}, to perform the utterance-level classification.  

\section{Emotional Corpora}
\label{sec:data}
Three different emotion corpora are used to evaluate the proposed method, namely, a Chinese emotion corpus (CASIA) \cite{tao2008design}, a German emotion corpus (Emo-DB) \cite{burkhardt2005database} and an English emotional database (SAVEE) \cite{jackson2014surrey}. They are summarized in Table~\ref{tab:data}. For each dataset, all of the emotion categories are selected for experiments.

Specifically, the CASIA corpus contains \(9,600\) speech utterances that were produced by four subjects (two males and two females) to stimulate six different emotions, i.\,e., anger, fear, happiness, neutral, sadness, and surprise. Only the \(7,200\) utterances corresponding to \(300\) linguistically neutral sentences with the same statements are involved in our experiments. The sample rate is \(16\) kHz.

The Berlin Emo-DB German Corpus (Emo-DB) contains \(535\) emotional speech utterances covering seven different emotions deliberately displayed by ten German actors (five males and five females). The number of spoken utterances for these seven emotions is not equally distributed: \(126\) anger, \(81\) boredom, \(46\) disgust, \(69\) fear, \(71\) happiness, \(79\) neutral, and \(62\) sadness. Audio files were recorded at \(16\) kHz. 

The Surrey audio-visual expressed emotion database (SAVEE) is a collection of read speech produced by four male British-English speakers (researchers and postgraduate students from the University of Surrey, age from \(27\) to \(31\) years). Seven different emotions are elicited: anger, disgust, fear, happiness, neutral, sadness, and surprise. The neutral emotion was uttered in \(30\) phonetically-balanced sentences, and each of the remaining emotions was expressed in \(15\) sentences. Sound files were sampled at \(44.1\) kHz.

\begin{table}[thbp]
  \centering
  \caption{The selected emotion corpora. (\# of Utt.: number of utterances; \# of Sub.: number of subjects; and \# of Emo.: number of emotions involved)}
  \resizebox{0.96 \linewidth}{!}{%
  \renewcommand\arraystretch{1.36}
  \begin{tabular}{p{1.1cm}p{1.1cm}<{\centering}p{1.1cm}<{\centering}p{1.6cm}<{\centering}p{1.2cm}<{\centering}p{1.6cm}<{\centering}}
  \toprule
   & Language & \# of Utt. & \# of Sub.  & \# of Emo. & Sampling rate \\
  \midrule
  \textbf{CASIA}  & Chinese & \(7,200\) & \(4\) (\(2\) female) & \(6\) & \(16\)kHz \\
  \textbf{Emo-DB}  & German & \(535\)  & \(10\) (\(5\) female) & \(7\) & \(16\)kHz \\
  \textbf{SAVEE}  & English & \(480\) & \(4\) (\(0\) female) & \(7\) & \(44.1\)kHz \\
  \bottomrule
  \end{tabular}%
  }
  \label{tab:data}
\end{table}



\begin{algorithm}[!tbp]
  \caption{Iterative Self-Learning Pseudocode}
  \begin{algorithmic}[ht!]
    \STATE {\bf Input:} speech segments with labels inherited from corresponding utterances (dataset with noisy labels)
    \STATE {\bf for} \(t = 1 : \textnormal{num\_iteration}\) \textbf{do}\\
    \quad update backbone DCNN parameter \(\theta_{t}\) by SGD \\
    \quad update labels by Eq. (\ref{eq:bdsl}) (BDSL)\\
    \quad\quad\quad\quad\quad\quad\,\,\, or Eq. (\ref{eq:wdsl}) (WDSL)\\
    \quad\quad\quad\quad\quad\quad\,\,\, or Eq. (\ref{eq:gsl}) (GSL)\\
    \quad\quad\quad\quad\quad\quad\,\,\, or Eq. (\ref{eq:dhl}) (DHL)\\
    \STATE {\bf end for}
    \STATE {\bf Output:} a robust backbone DCNN to make more corrected segment-level predictions, which will be used for constructing utterance-level representations
  \end{algorithmic}
  \label{algo: sl}
\end{algorithm}

\section{Acoustic Features}
In regards to the acoustic features, we follow the recent success of applying DCNN directly to (Mel) spectrograms \cite{fayek2017evaluating, satt2017efficient, ma2018emotion, mao2019deep}. Empirical comparisons between automatic features learned from spectrograms and standard human-engineered features in our prior work \cite{mao2019deep} have demonstrated the superiority of the former. 

\begin{table*}[t]
  \caption{Results of the DSL-BDSL method across various backbone network architectures on CASIA corpus, Emo-DB database and SAVEE database, respectively. In each subtable, the first model was trained using the original noisy labels, and each subsequent model was trained using labels generated by the model immediately above it; the superscript of each model stands for the iteration number of self-learning. Recognition rate is reported in [\%], and the highest one is marked in bold.}
  \begin{minipage}{0.48\linewidth}
  \centering
  \resizebox{1. \linewidth}{!}{%
  \begin{tabular}{p{1.8cm}|c|c|c|c|c|c}
  \toprule
  \multicolumn{1}{c|}{} & \multicolumn{2}{c|}{CASIA} & \multicolumn{2}{c|}{Emo-DB} &\multicolumn{2}{c}{SAVEE} \\
  \midrule
  \bf{Model} & WA & UA & WA & UA & WA & UA\\
  \midrule
  \(\textnormal{VGG}19^0\) & $93.10$ & $93.10$ & $83.00$ & $82.36$ & $70.63$ & $69.88$\\
  \rowcolor[gray]{.8} \(\textnormal{VGG}19^1\) & $\bf93.67$ & $\bf93.67$ & $\bf83.74$  & $\bf83.96$ & $\bf71.88$ & $\bf70.00$\\
  \(\textnormal{VGG}19^2\) & $90.07$ & $90.07$ & $69.91$ & $67.92$ & $26.04$ & $21.07$\\
  \bottomrule
  \end{tabular}
  }
  \end{minipage}
  \hspace{5mm}
  \begin{minipage}[r]{0.48\linewidth}
  \centering
  \resizebox{1. \linewidth}{!}{%
  \begin{tabular}{p{1.8cm}|c|c|c|c|c|c}
  \toprule
  \multicolumn{1}{c|}{} & \multicolumn{2}{c|}{CASIA} & \multicolumn{2}{c|}{Emo-DB} &\multicolumn{2}{c}{SAVEE} \\
  \midrule
  \bf{Model} & WA & UA & WA & UA & WA & UA\\
  \midrule
  \(\textnormal{DenseNet}22^0\) & $87.42$ & $87.42$ & $80.37$ & $79.99$ & $63.75$ & $60.60$\\
  \rowcolor[gray]{.8} \(\textnormal{DenseNet}22^1\) & $\bf91.11$ & $\bf91.11$ & $\bf81.68$ & $\bf81.41$ & $\bf64.79$ & $\bf61.31$\\
  \(\textnormal{DenseNet}22^2\) & $70.79$ & $70.79$ & $67.23$ & $67.18$ & $43.13$ & $38.10$\\
  \bottomrule
  \end{tabular}
  }
  \end{minipage}
  \label{tab:soft_no_hard}
  \end{table*}

\begin{table*}[t]
  \begin{minipage}{0.48\linewidth}
    \centering
    \resizebox{1. \linewidth}{!}{%
    \begin{tabular}{p{1.8cm}|c|c|c|c|c|c}
    \toprule
    \multicolumn{1}{c|}{} & \multicolumn{2}{c|}{CASIA} & \multicolumn{2}{c|}{Emo-DB} &\multicolumn{2}{c}{SAVEE} \\
    \midrule
    \bf{Model} & WA & UA & WA & UA & WA & UA\\
    \midrule
    \(\textnormal{MobileNetV}2^0\) & $95.01$ & $95.01$ & $81.31$ & $81.05$ & $72.92$ & $70.60$\\
    \rowcolor[gray]{.8} \(\textnormal{MobileNetV}2^1\) & $\bf95.76$ & $\bf95.76$ & $\bf89.53$ & $\bf89.21$ & $\bf73.29$ & $\bf71.88$\\
    \(\textnormal{MobileNetV}2^2\) & $93.15$ & $93.15$ & $68.22$ & $65.51$ & $26.67$ & $21.07$\\
    \bottomrule
    \end{tabular}
    }
  \end{minipage}
  \hspace{5mm}
  \begin{minipage}[r]{0.48\linewidth}
    \centering
    \resizebox{1. \linewidth}{!}{%
    \begin{tabular}{p{1.8cm}|c|c|c|c|c|c}
    \toprule
    \multicolumn{1}{c|}{} & \multicolumn{2}{c|}{CASIA} & \multicolumn{2}{c|}{Emo-DB} &\multicolumn{2}{c}{SAVEE} \\
    \midrule
    \bf{Model} & WA & UA & WA & UA & WA & UA\\
    \midrule
    \(\textnormal{EfficientNet-B0}^0\) & $95.17$ & $95.17$ & $83.36$ & $82.54$ & $73.75$ & $72.26$\\
    \rowcolor[gray]{.8} \(\textnormal{EfficientNet-B0}^1\) & $\bf95.61$ & $\bf95.61$ & $\bf91.78$  & $\bf91.37$ & $\bf84.79$ & $\bf83.93$\\
    \(\textnormal{EfficientNet-B0}^2\) & $92.85$ & $92.85$ & $85.42$ & $85.15$ & $80.63$ & $79.05$\\
    \bottomrule
    \end{tabular}
    }
  \end{minipage}
\end{table*}

Specifically, our backbone DCNN works on the \(64\)-bin log Mel filterbanks of individual segments. To calculate this spectrogram features for the segments, we first resample the speech signals in the SAVEE database to \(16\)kHz using the Librosa framework \cite{mcfee2015librosa}, such that all audio files of the three emotional corpora have the same sampling rate. Then, a sequence of overlapping Hanning windows is applied to the speech signals, resulting in frames with window shift of \(10\) ms, and window size of \(25\) ms. Subsequently, for each frame, the STFT is computed with an FFT length of \(512\) points. Finally, we compute the logarithmic power of \(64\) Mel-frequency filterbanks over a frequency range from \(0.125\) kHz to \(7.5\) kHz, where the formula for converting from frequency \(f\) to Mel scale \(M(f)\) is:
\begin{align}
  M(f) = 1127\,\textnormal{ln}\left(1 + \frac{f}{700}\right)
  \label{mel}
\end{align}

The resulting frame-level log Mel filterbanks are then concatenated to form a 2-dim time-frequency representation of the segment. In this work, the segment size is set to \(32\) frames, i.\,e., the total length of a segment is \(10\) ms \(\times\) \(32\) \(+\) (\(25\) - \(10\)) ms = \(335\) ms. For the CASIA corpus, the segment hop length is set to \(30\) ms, while it is set to \(10\) ms for the remaining two database. In this way, we collected \(418,722\) segments for the CASIA corpus, \(131,053\) segments for the Emo-DB database, and \(51,027\) segments for the SAVEE database, for training the backbone network, respectively. We also attempted other segment sizes ranging from \(215\) ms to \(415\) ms with the same hop length and achieved similar performance for utterance-level classification.

\section{Backbone Networks}
Deep convolutional neural networks (DCNN) have proven very effective in image classification and show promise for speech signals. In this work, we have investigated various DCNN architectures as the backbone network to generate segment-level emotion predictions. Specifically, we examined VGG\(19\) \cite{simonyan2014very}, DenseNet\(22\) \cite{huang2017densely}, MobilenetV\(2\) \cite{sandler2018mobilenetv2}, and EfficientNet-B0 \cite{tan2019efficientnet}. For each of the DCNNs, the architecture of the convolutional layers is based on the configurations in the original paper. A slight adjustment is made to the neuron number in the last softmax layer in order to make it suitable for our tasks. Note that we did not put much effort into optimization of the network architectures since our major concern here is the DSL framework. 

\section{Experiments}
We evaluate the proposed method on the three mentioned emotion corpora. We first explore the effect of BDSL, which updates the training targets without combining the initial noisy label. We then evaluate the WDSL method, which remains the initial noisy label as a part of supervision. Subsequently, we evaluate the performance of the GSL and HDL methods, where we also gain some insight into the source of the improvements obtained using the BDSL and WDSL methods. Finally, we present some ablation studies and analyses to investigate the effect of the weight coefficient \(\alpha\) in the WDSL method. The WDSL method with \(\alpha\) equal to \(0.2\) achieves the best results across the three emotion corpora.

\subsection{Setup}
For the DCNN training, the ADAM \cite{kingma2014adam} optimizer with default setting in Tensorflow \cite{abadi2016tensorflow} is used. The initial learning rate is set to \(0.001\), and an exponential decay scheme with a rate of \(0.8\) every two epochs is applied. The batch size is set to \(128\). Early stopping strategy with patience of \(3\) epochs is applied to mitigate an overfitting problem. Maximum number of training epochs is set to \(20\) for the CASIA corpus, \(12\) for the Emo-DB corpus, and eight for the SAVEE database. 

The segment-level predictions are generated using \(10\)-fold stratified cross-validation to ensure that the segment-level predictions are out-of-sample. Furthermore, the fold split is done at the utterance level and not at the segment level. The utterance-level classification is performed using a random forest (RF) with the default setting in the open-source Scikit-learn toolkit \cite{pedregosa2011scikit}, where another \(10\)-fold stratified cross-validation is performed. The results are presented in terms of weighted accuracy (WA) and unweighted accuracy (UA). Note that the WA and UA are the same for the CASIA corpus, as the CASIA corpus is perfectly balanced regarding the emotion class distribution.

\subsection{Results of DSL-BDSL}
We first investigate the effect of applying DSL with basic dynamic soft labeling (DSL-BDSL), in which we use the soft labels (model predictions) solely as the supervision signal for training next model, without combining the initial noisy label. Table \ref{tab:soft_no_hard} shows the experimental results for DSL-BDSL on various DCNN architectures. In each sub-table, the superscript of each model stands for the iteration number of self-learning; each row represents a randomly-initialized model trained with soft labels generated by the model one row above it. For instance, \(\textnormal{VGG}19^2\) is trained with soft labels generated by \(\textnormal{VGG}19^1\), and \(\textnormal{VGG}19^1\) is trained with class probabilities predicted by \(\textnormal{VGG}19^0\). The first model \(\textnormal{VGG}19^0\) is trained with the original predefined noisy labels.

As can be observed from Table \ref{tab:soft_no_hard}: (1) The DSL-BDSL method consistently improved the SER accuracy on all three emotional corpora across all four DCNN architectures, demonstrating the universality of the proposed method. (2) The best performance was achieved after one single round of re-training process. Subsequently, the performance diminished significantly. (3) The best performance varied with the backbone DCNN architectures. In particular, the recognition rate on the Emo-DB corpus of VGG\(19\) (\(83.74\%\) for WA and \(83.96\%\) for UA) and DenseNet\(22\) (\(81.68\%\) for WA and \(81.41\%\) for UA) were noticeably lower than that of MobileNetV\(2\) (\(89.53\%\) for WA and \(89.21\%\) for UA) and EfficientNet-B0 (\(91.78\%\) for WA; \(91.37\%\) for UA). The latter two architectures use the \textit{mobile inverted bottleneck} \cite{sandler2018mobilenetv2, tan2019mnasnet, tan2019efficientnet} as their main building block, in which the lightweight design of depthwise separable convolutions \cite{sandler2018mobilenetv2, tan2019mnasnet, tan2019efficientnet} might offer an advantage in further mitigating the data scarcity problem. Moreover, the overal superiority of EfficientNet-B0 over MobileNetV\(2\) can be attributed to that the former was developed with a greedy neural architecture search (NAS) algorithm \cite{zophneural, tan2019efficientnet} and thus better performance can be expected. Consequently, EfficientNet-B0 will be used in the remaining experiments.

Figure~\ref{fig:bdsl} shows an example of segment-level label predictions across the audio file ``Happy\_liuchanhg\_382.wav'' in the CASIA corpus, where the DSL-BDSL method was applied, and EfficientNet-B0 was used as the backbone network. It is obvious that the DSL-BDSL method encourages the predicted class probabilities to be near a uniform distribution, i.\,e., after the second iteration of re-training, each output dimension of the backbone network was already fairly close to \(17\%\), i.\,e., the chances of a random hit for the CASIA corpus of six categories. We argue that this is because, for the DSL-BDSL method, the model minimizes Eq. (\ref{eq:entropy-bdsl}) solely. As the self-learning process continues, we gradually lost useful information in the original noisy one-hot label, which encourages sharp \(1\)-of-\(K\) (one-hot) code predictions. Finally, we arrived at a trivial global optimal solution for Eq. (\ref{eq:entropy-bdsl}), where a backbone model always makes predictions to be near a uniform distribution. 

\subsection{Results of DSL-HDL and DSL-GSL}

\begin{table}[!tbp]
  \renewcommand\arraystretch{1}
  \caption{Results of the DSL-GSL and DSL-HDL method on CASIA corpus, Emo-DB database and SAVEE database, respectively. For comparison, we also list the results obtained using the DSL-BDSL method as well as the baseline in which no DSL was applied. EfficientNet-B0 was used as the backbone model. Recognition rate is reported in [\%], and the highest one is marked in bold.}
  \resizebox{1. \linewidth}{!}{%
  \begin{tabular}{p{1.8cm}|c|c|c|c|c|c}
  \toprule
  \multicolumn{1}{c|}{} & \multicolumn{2}{c|}{CASIA} & \multicolumn{2}{c|}{Emo-DB} &\multicolumn{2}{c}{SAVEE} \\
  \midrule
  \bf Method & WA & UA & WA & UA & WA & UA\\
  \midrule
  No DSL & $95.17$ & $95.17$ & $83.36$ & $82.54$ & $73.75$ & $72.26$ \\
  DSL-GSL & $95.25$  & $95.25$  & $84.11$ & $84.15$ & $76.25$ & $74.76$ \\
  DSL-HDL & $95.56$ & $95.56$ & $91.03$ & $90.85$ & $81.25$ & $79.76$\\
  \rowcolor[gray]{.8} DSL-BDSL & $\bf95.61$ & $\bf95.61$ & $\bf91.78$  & $\bf91.37$ & $\bf84.79$ & $\bf83.93$\\
  \bottomrule
  \end{tabular}%
  }
  \label{tab:gsl_n_hdl}
\end{table}

\begin{figure}[t]
  \centering
  \includegraphics[width=\linewidth]{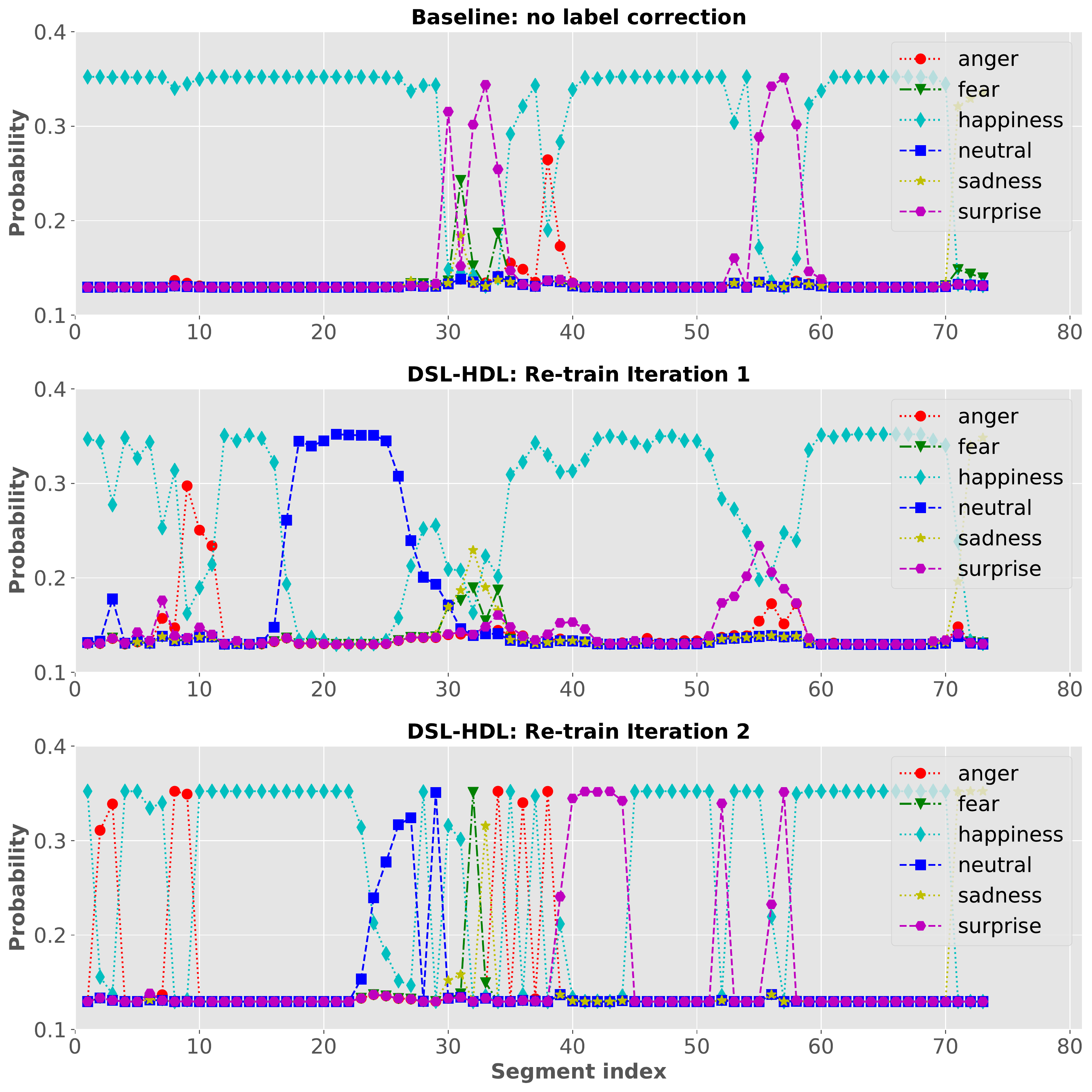}
  \caption{An example of segment-level label predictions across the audio file ``Happy\_liuchanhg\_382.wav'' in the CASIA corpus. The DSL-HDL method was applied, and EfficientNet-B0 was used as the backbone network.}
  \label{fig:hdl}
\end{figure}

We posit that the benefits of using DSL-BDSL are from two aspects: (1) Each segment is dynamically re-labeled with a more appropriate label; and (2) soft labeling is introduced. To observe the improvement from dynamic labeling alone, we perform label refinement with hard dynamic labeling (HDL) in the DSL framework, which is called DSL-HDL. Specifically, we pass each segment to the trained backbone model, and the one-hot label is assigned by choosing the most-likely category from the model output. Figure~\ref{fig:hdl} shows the segment-level label predictions across the audio file ``Happy\_liuchanhg\_382.wav'' in the CASIA corpus, with the DSL-HDL method applied and EfficientNet-B0 as the backbone. It can be observed that, as the self-training proceeds, the emotion prediction contour across the utterance oscillated in an increasingly strong manner.

\begin{figure}[t]
  \centering
  \includegraphics[width=\linewidth]{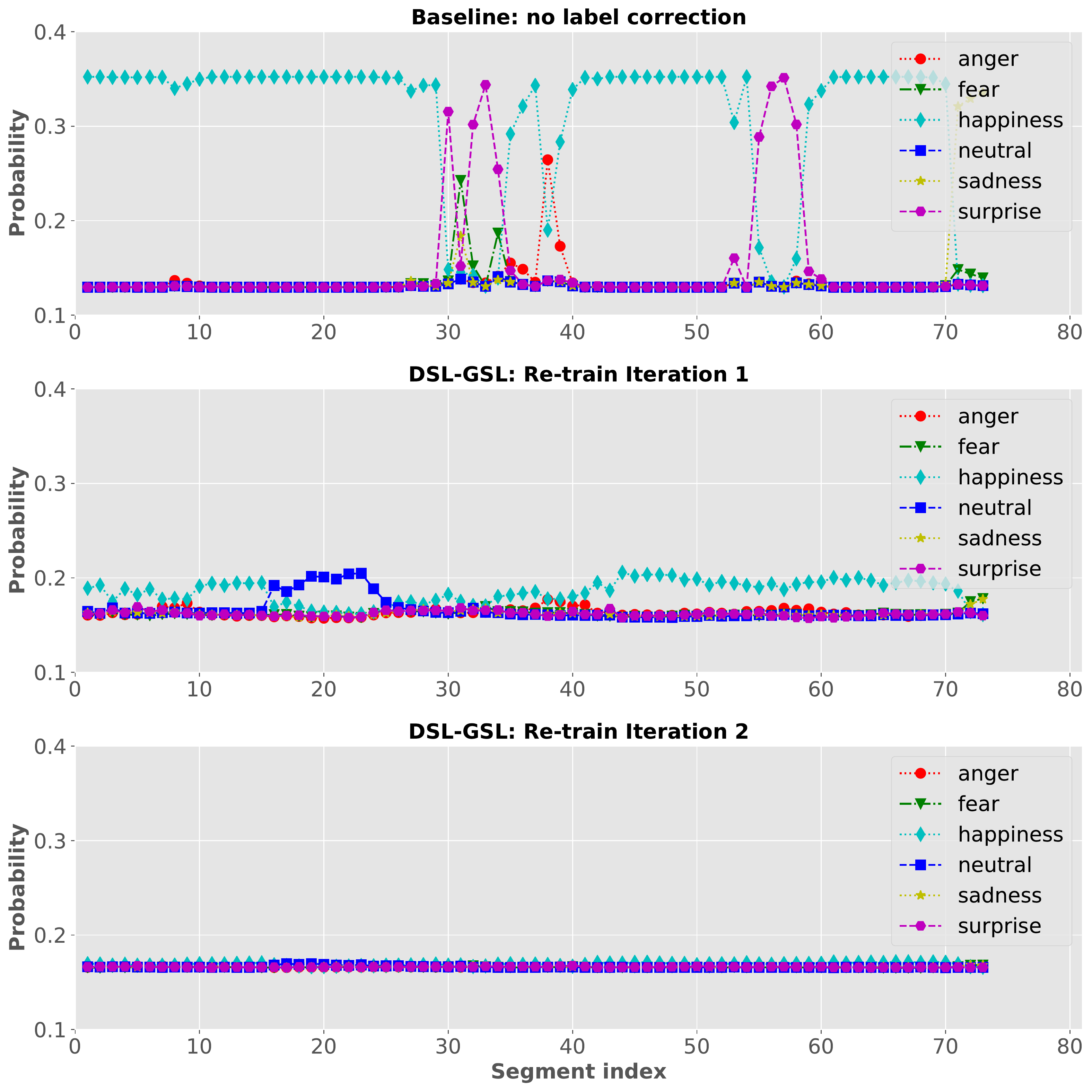}
  \caption{An example of segment-level label predictions across the audio file ``Happy\_liuchanhg\_382.wav'' in the CASIA corpus. The DSL-GSL method was applied, and EfficientNet-B0 was used as the backbone network.}
  \label{fig:gsl}
\end{figure}

To assess the improvement from soft labeling solely, we investigate global soft labeling (GSL) in the DSL framework, which is called DSL-GSL. To attain the soft global label for a given segment, we pass all segments within the same utterance to the trained backbone model, and the soft global label is calculated by averaging the model outputs over the whole utterance. Figure~\ref{fig:gsl} presents the segment-level label predictions across the audio file ``Happy\_liuchanhg\_382.wav'' in the CASIA corpus, with the DSL-GSL method applied and EfficientNet-B0 as the backbone. Similar to the DSL-BDSL method, after two iterations of self-learning, the model in the case of using the DSL-GSL method also arrives at a state where it always makes uniformly distributed predictions. 

Table~\ref{tab:gsl_n_hdl} summarizes the results. Both DSL-HDL and DSL-GSL method consistently boost the recognition accuracy. When they are combined (i.\,e., the DSL-BDSL method), an additional improvement can be observed, indicating that they address different issues in labels.

\subsection{Results of DSL-WDSL}

\begin{figure}[t]
  \centering
  \includegraphics[width=\linewidth]{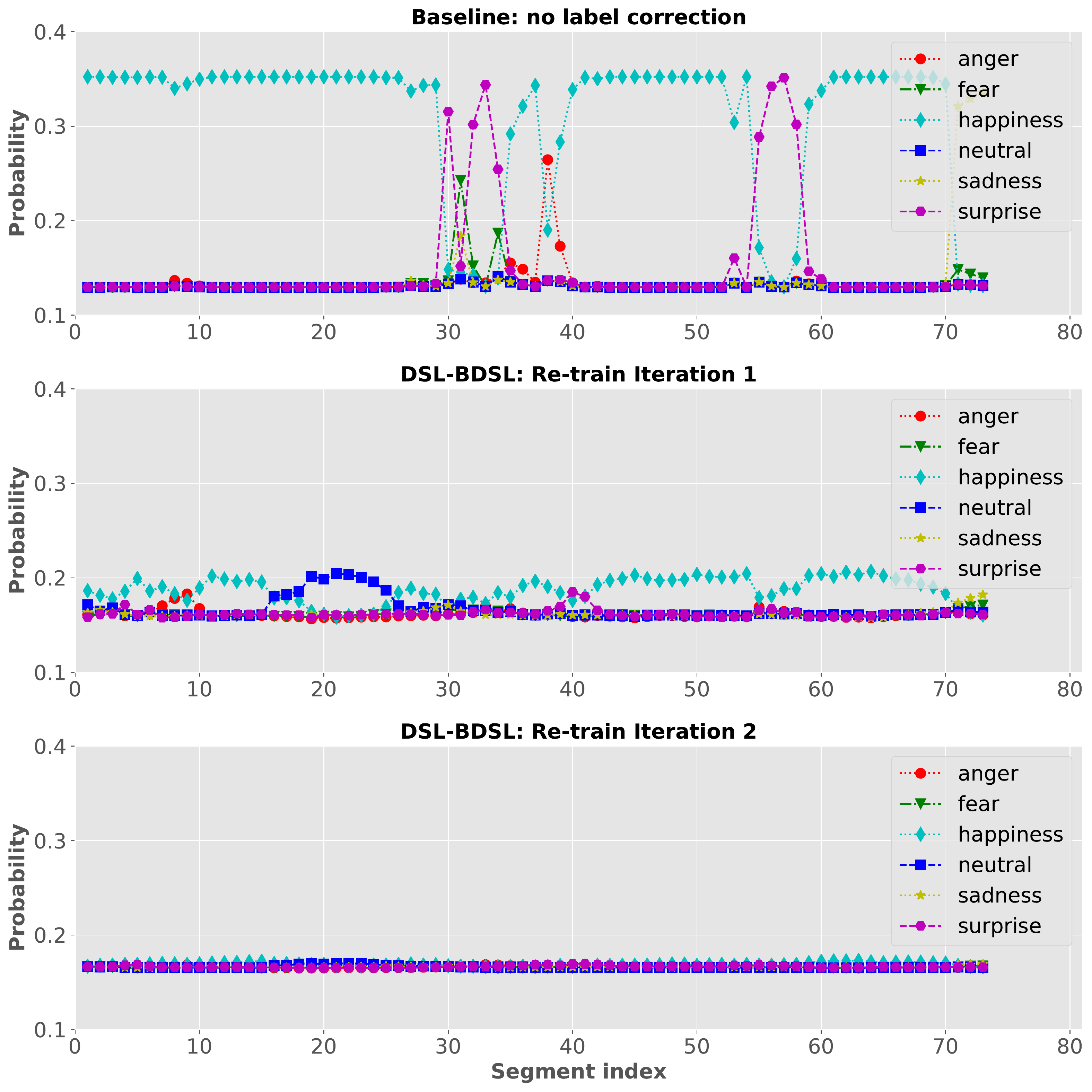}
  \caption{An example of segment-level label predictions across the audio file ``Happy\_liuchanhg\_382.wav'' in the CASIA corpus. The DSL-BDSL method was applied, and EfficientNet-B0 was used as the backbone network.}
  \label{fig:bdsl}
\end{figure}
\begin{figure}[t]
  \centering
  \includegraphics[width=\linewidth]{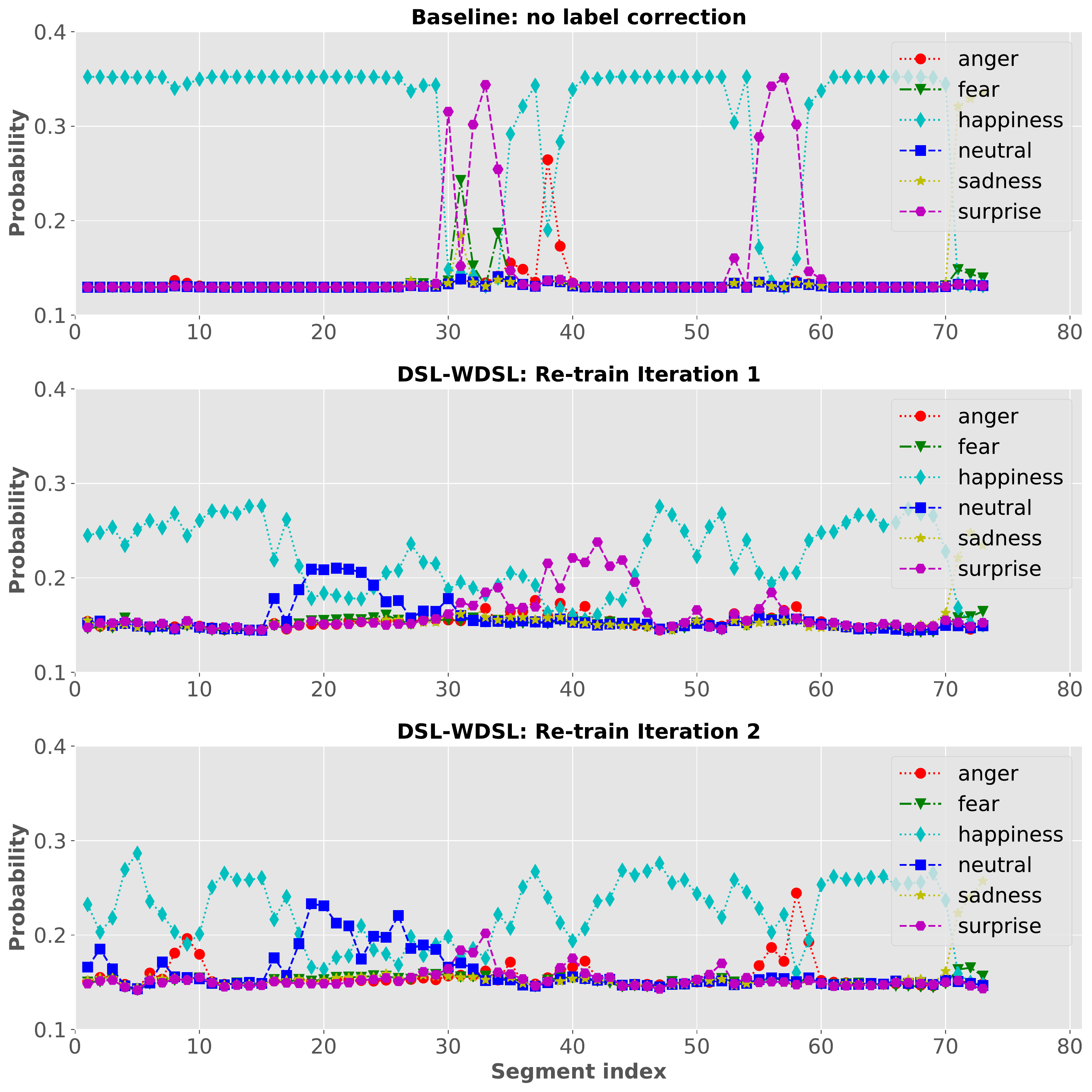}
  \caption{An example of segment-level label predictions across the audio file ``Happy\_liuchanhg\_382.wav'' in the CASIA corpus. The DSL-WDSL method was applied, and EfficientNet-B0 was used as the backbone network.}
  \label{fig:wdsl}
\end{figure}

In the previous DSL-BDSL method, as self-learning proceeded to the second iteration, we got stuck in a trivial solution for Eq. (\ref{eq:entropy-bdsl}), where the segment-level model always made predictions to be near a uniform distribution. In this section, we tackle this issue by keeping the original noisy label as a part of the supervision. Eq. (\ref{eq:entropy-wdsl}) shows the resulting loss function, where the second term (introduced by the original noisy label) is equivalent to the conventional softmax regression, which avoids the trivial global minima of Eq. (\ref{eq:entropy-bdsl}) by encouraging \(1\)-of-\(K\) one-hot code predictions, as well as adding a strong bias towards the corresponding utterance-level emotion class (as it should be). Note that there is a weight coefficient \(\alpha\) that balances the two terms in Eq. (\ref{eq:entropy-wdsl}). In the very beginning, we heuristically set \(\alpha\) equal to \(0.5\). An ablation study on the influence of different \(\alpha\) will be presented in the following section. Figure~\ref{fig:wdsl} shows the segment-level label predictions across the audio file ``Happy\_liuchanhg\_382.wav'' in the CASIA corpus, with the DSL-WDSL method applied and EfficientNet-B0 as the backbone. It can be observed that the uniformly distributed predictions that were encountered in the case of using DSL-BDSL (see Figure~\ref{fig:bdsl}) disappeared, indicating the promising potential of combining both the soft labels and original noisy labels in the DSL scheme.

\begin{figure*}[!htbp]
  \centering
  \includegraphics[width=0.93\linewidth]{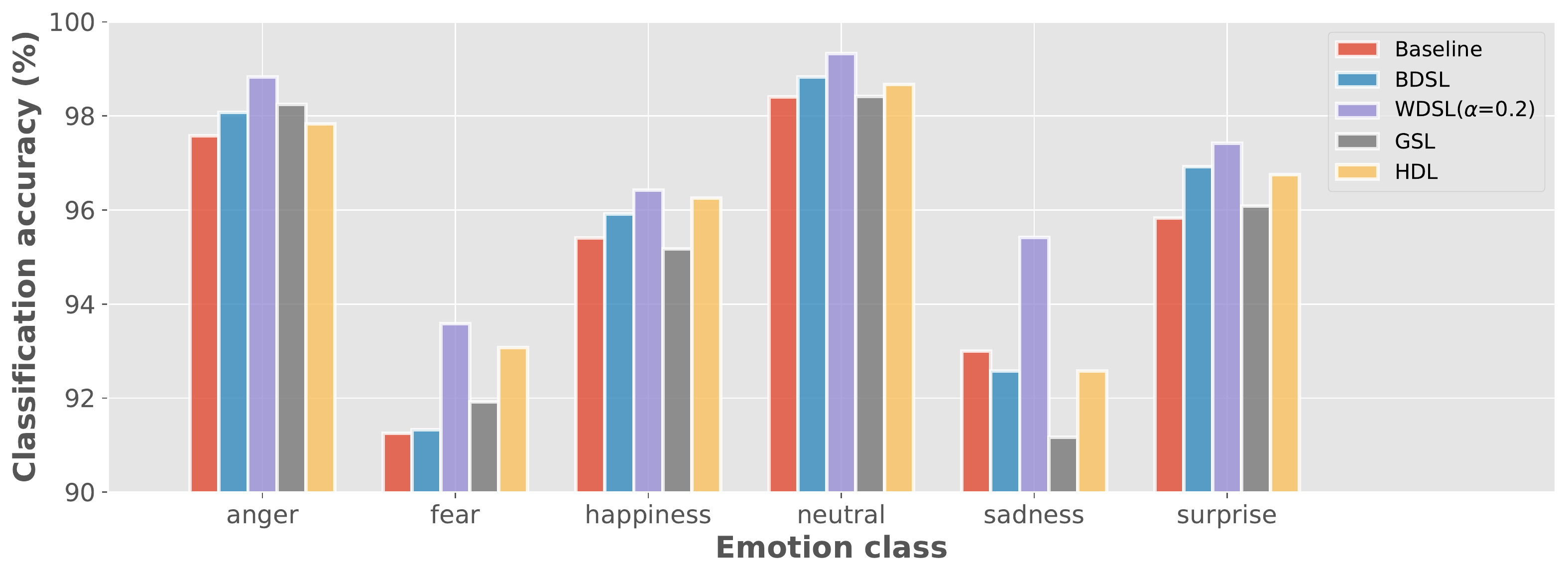}
  \caption{Utterance-level accuracy [\%] for each class in the CASIA corpus obtained using: (1) original noisy segment labels, without DSL method applied (Baseline); (2) labels corrected by the DSL-BDSL method (BDSL); (3) labels corrected by the DSL-WDSL method, where the weight factor \(\alpha = 0.2\) (WDSL (\(\alpha = 0.2\))); (4) labels corrected by the DSL-GSL method (GSL); (5) labels corrected by the DSL-HDL method (HDL), respectively.}
  \label{fig:casia_acc}
\end{figure*}

\begin{figure*}[!htbp]
  \centering
  \includegraphics[width=0.93\linewidth]{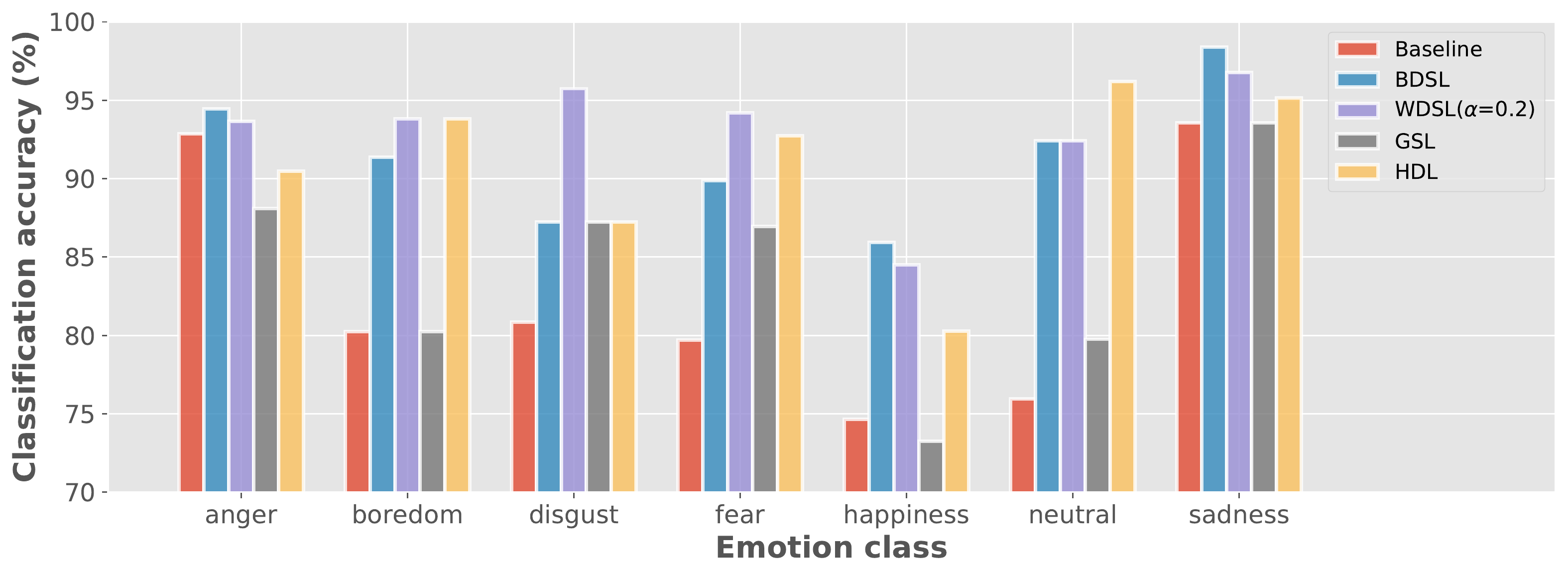}
  \caption{Utterance-level accuracy [\%] for each class in the Emo-DB corpus obtained using: (1) original noisy segment labels, without DSL method applied (Baseline); (2) labels corrected by the DSL-BDSL method (BDSL); (3) labels corrected by the DSL-WDSL method, where the weight factor \(\alpha = 0.2\) (WDSL (\(\alpha = 0.2\))); (4) labels corrected by the DSL-GSL method (GSL); (5) labels corrected by the DSL-HDL method (HDL), respectively.}
  \label{fig:emodb_acc}
\end{figure*}

\begin{figure*}[!htbp]
  \centering
  \includegraphics[width=0.93\linewidth]{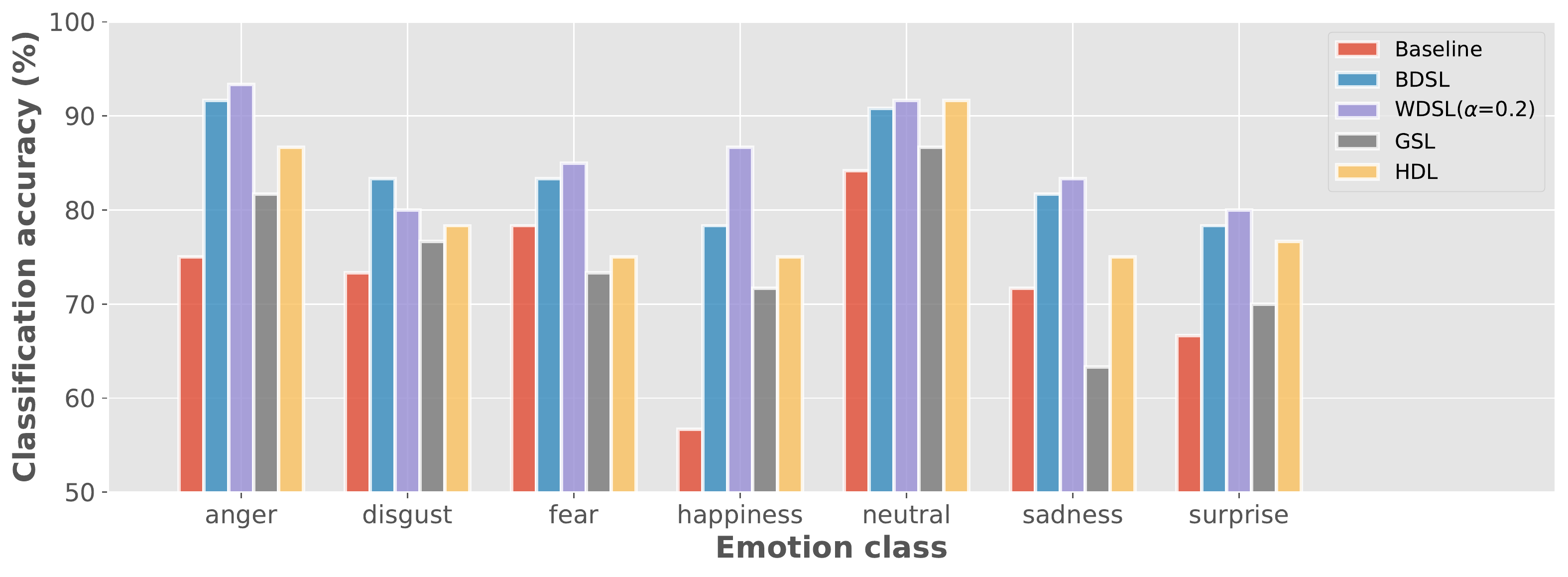}
  \caption{Utterance-level accuracy [\%] for each class in the SAVEE database obtained using: (1) original noisy segment labels, without DSL method applied (Baseline); (2) labels corrected by the DSL-BDSL method (BDSL); (3) labels corrected by the DSL-WDSL method, where the weight factor \(\alpha = 0.2\) (WDSL (\(\alpha = 0.2\))); (4) labels corrected by the DSL-GSL method (GSL); (5) labels corrected by the DSL-HDL method (HDL), respectively.}
  \label{fig:savee_acc}
\end{figure*}

\subsection{Ablation Study on the Weight Factor \(\alpha\)}

The weight factor \(\alpha\) in Eq.~(\ref{eq:entropy-wdsl}) is crucial for the DSL-WDSL method. If \(\alpha\) is too large, the model concentrates on the initial noisy labels, and it disturbs the training. On the other hand, if \(\alpha\) is too small, the training procedure discards the useful information in the initial noisy labels, leading to the trivial global optima as encountered in the previous DSL-BDSL method. In this section, we study the influence of different \(\alpha\). Specifically, \(\alpha\) is constant ranging from \(0.0\) to \(0.9\). In addition, we also investigated the deterministic annealing process as used in \cite{lee2013pseudo}, by which \(\alpha\) is slowly decreased (i.\,e., \(0.9 \rightarrow 0.7 \dots \rightarrow 0.1\)), but the improvement in our experiment was somewhat limited. 

Table~\ref{tab:alpha} presents the results, where EfficientNet-B0 was applied as the backbone. We find that when the DSL method was not applied, which corresponds to \(\alpha = 1\) in the last row, the performance was relatively limited. On the other hand, the model trained using the generated label solely (i.\,e., the DSL-BDSL method that corresponds to \(\alpha = 0\) in Table~\ref{tab:alpha}) also performed sub-optimally. When \(\alpha = 0.2\), the model jointly trained using the original noisy label and the generated soft label achieved the best performance across all three databases.

\begin{table}[!tbp]
  \renewcommand\arraystretch{1}
  \centering
  \caption{Results of the DSL-WDSL method on CASIA corpus, Emo-DB database and SAVEE database, respectively. The weight coefficient \(\alpha\) ranges from \(0\) to \(1\). Note that \(\alpha = 0\) is equivalent to applying DSL-BDSL method, and \(\alpha = 1\) falls back to the case where no DSL is applied. EfficientNet-B0 was used as the backbone. Recognition rate is reported in [\%], and the highest one is marked in bold.}
  \resizebox{0.946 \linewidth}{!}{%
  \begin{tabular}{c|c|c|c|c|c|c}
  \toprule
  \multicolumn{1}{c|}{} & \multicolumn{2}{c|}{CASIA} & \multicolumn{2}{c|}{Emo-DB} &\multicolumn{2}{c}{SAVEE} \\
  \midrule
  $\bf{\alpha}$ & WA & UA & WA & UA & WA & UA\\
  \midrule
  $0$ & $95.61$ & $95.61$ & $91.78$ & $91.37$ & $84.79$ & $ 83.93$\\
  $0.1$ & $96.14$ & $96.14$ & $\bf92.90$ & $92.99$ & $84.38$ & $82.74$\\
  \rowcolor[gray]{.8} $0.2$ & $\bf96.82$ & $\bf96.82$ & $\bf92.90$ & $\bf93.02$ & $\bf86.46$ & $\bf85.71$\\
  $0.3$ & $96.76$ & $96.76$ & $92.15$ & $92.12$ & $81.67$ & $80.12$\\
  $0.4$ & $96.79$ & $96.79$ & $91.78$ & $91.62$ & $83.75$ & $82.50$\\
  $0.5$ & $96.67$ & $96.67$ & $90.65$ & $90.64$ & $82.29$ & $80.71$\\
  $0.6$ & $96.57$ & $96.57$ & $90.09$ & $89.72$ & $81.25$ & $79.29$\\
  $0.7$ & $96.47$ & $96.47$ & $88.97$ & $89.03$ & $81.25$ & $80.24$\\
  $0.8$ & $96.33$ & $96.33$ & $88.60$ & $88.49$ & $79.17$ & $77.50$\\
  $0.9$ & $96.24$ & $96.24$ & $88.41$ & $88.19$ & $77.92$ & $76.55$\\
  \midrule
  \midrule
  $1$ & $95.17$ & $95.17$ & $83.36$ & $82.54$ & $73.75$ & $72.26$\\
  \bottomrule
  \end{tabular}%
  }
  \label{tab:alpha}
\end{table}

\subsection{Discussion}
Table~\ref{tab:compare} summarizes the results on the three mentioned emotional corpora, respectively, where EfficientNet-B0 was used as the backbone. The following can be seen: (1) Our baseline system achieved recognition rate of \(95.17\%\) for the CASIA corpus, which already surpasses the state-of-the-art performance in the literature, demonstrating the effectiveness of the segment-based approach. On the other hand, our baseline system did not show any superiority on the two smaller datasets, i.\,e., the Emo-DB corpus and the SAVEE database. The main reason for this is that, in the segmentation process, the segment hop length was only set to \(10\) ms for the Emo-DB corpus and the SAVEE database, whereas it was set to \(30\) ms for the CASIA corpus. Indeed, a smaller hop length (equivalent to larger overlapping) yields more segments for a given utterance, but it also results in more severe homogeneity among segments for training. This trade-off is more difficult for smaller datasets. (2) The proposed DSL framework consistently improved the baseline system, providing an effective and efficient scheme for training a robust emotion segment model on noisy-labeled speech segments. (3) All of the dynamic-based DSL methods including DSL-BDSL, DSL-WDSL and DSL-HDL consistently outperformed the static-based DSL-GSL method, indicating that labels should be defined at the segment-level rather than being identical across the whole utterance, which partly justified our motivation for this work. (4) The DSL-WDSL method substantially augmented the performance of the DSL-BDSL method, verifying that original noisy labels are useful in the training procedure. To our best knowledge, the DSL-WDSL method established new state-of-the-art performance on all three emotional corpora. (5) Figures~\ref{fig:casia_acc}-\ref{fig:savee_acc} show the utterance-level recognition accuracy for different classes across the three emotional corpora, respectively. The proposed DSL framework improved classification accuracy for most emotion classes.

\begin{table}[!tbp]
  \renewcommand\arraystretch{1}
  \centering
  \caption{Results of different DSL-based methods on CASIA corpus, Emo-DB database and SAVEE database, respectively. The weight coefficient \(\alpha\) in the DSL-WDSL method is set to \(0.2\). EfficientNet-B0 was used as the backbone. Recognition rate is reported in [\%], and the highest one is marked in bold. Note that the WA and UA are the same for the CASIA corpus, we only list WA here for the CASIA corpus.}
  \resizebox{1.0 \linewidth}{!}{%
  \begin{tabular}{l|c|c|c|c|c}
  \toprule
  \multicolumn{1}{c|}{} & \multicolumn{1}{c|}{CASIA} & \multicolumn{2}{c|}{Emo-DB} &\multicolumn{2}{c}{SAVEE} \\
  \midrule
  $\bf{Methods}$ & WA & WA & UA & WA & UA\\
  \midrule
  Li et al. \cite{li2015simulated} & $80.60$ & $84.80$ & $\bf{-}$ & $65.00$ & $\bf{-}$\\
  Sun et al. \cite{sun2015weighted} & $85.08$ & $89.32$ & $\bf{-}$ & $75.60$ & $\bf{-}$\\
  Sun et al. \cite{sun2017ensemble} & $\bf{-}$ & $88.70$ & $87.90$ & $76.30$ & $73.40$\\
  Liu et al. \cite{liu2018speech} & $90.28$ & $\bf{-}$ & $\bf{-}$ & $76.40$ & $\bf{-}$\\
  Mao et al. \cite{mao2018effective} & $90.90$ & $89.40$ & $87.88$ & $76.20$ & $76.20$\\
  Mao et al. \cite{mao2019revisiting} & $91.32$ & $90.48$ & $88.25$ & $\bf{-}$ & $\bf{-}$\\
  \midrule
  NO DSL (baseline) & $95.17$ & $83.36$ & $82.54$ & $73.75$ & $72.26$\\
  \midrule
  DSL-BDSL (ours) & $96.67$ & $91.78$  & $91.37$ & $84.79$ & $83.93$\\
  \rowcolor[gray]{.8} DSL-WDSL (ours) & $\bf96.82$ & $\bf92.90$ & $\bf93.02$ & $\bf86.46$ & $\bf85.71$\\
  DSL-GSL (ours) & $95.25$ & $84.11$ & $84.15$ & $76.25$ & $74.76$ \\
  DSL-HDL (ours) & $95.56$ & $91.03$ & $90.85$ & $81.25$ & $79.76$\\
  \bottomrule
  \end{tabular}%
  }
  \label{tab:compare}
\end{table}

\section{Conclusions}
In this paper, we address shortcomings commonly found in segment-based speech emotion recognition (SER). We treat the original segment-level labels that merely inherited from corresponding utterances as noisy (incorrect) labels and formulate the corresponding emotion segment modeling as the ``learning with noisy labels'' problem. We propose a deep self-learning (DSL) framework to progressively update the labels for network re-training to improve the emotion segment model's robustness. Also, considering the impurity nature of emotional speech, the soft labeling approach is investigated within the DSL framework, which characterizes the underlying mixture of emotions by representing each segment label with an emotion class distribution. We achieve a noticeable gain in emotion recognition performance across a broad range of network architectures, including VGG\(19\), DenseNet\(22\), MobileNetV\(2\), and EfficientNet-B0. Our experimental evaluation shows substantial improvement in the state-of-the-art accuracy on three well-known emotional corpora, respectively. In the future, we will explore combining self-learning with contrastive learning further to boost the generalization performance of the emotion segment model. Optimization of the network architectures will also be addressed.

\newpage

\bibliographystyle{IEEEtran}
\bibliography{mybib}


\end{document}